\documentclass[a4paper,aps,prl,showpacs,amssymb,twocolumn,superscriptaddress]{revtex4}
\usepackage{graphicx}

\begin{document}
%\draft
\date{\today}

\title{Stochastic resonance in a model of opinion formation on
small-world networks}

\author{Marcelo Kuperman}
\email{kuperman@cab.cnea.gov.ar} \affiliation{Centro At{\'o}mico
Bariloche and Instituto Balseiro, 8400 S. C. de Bariloche,
Argentina}
\author{Dami\'an Zanette}
\email{zanette@cab.cnea.gov.ar} \affiliation{Centro At{\'o}mico
Bariloche and Instituto Balseiro, 8400 S. C. de Bariloche,
Argentina} \affiliation{Consejo Nacional de Investigaciones
Cient{\'\i}ficas y T{\'e}cnicas, Argentina}

\begin{abstract}
{\normalsize We analyze  the phenomenon  of stochastic  resonance
in  an Ising-like system on a small-world network.  The system,
which is subject to  the combined  action  of  noise  and   an
external  modulation,  can   be interpreted  as  a  stylized
model  of opinion formation by imitation under the effects of a
``fashion wave''. Both the amplitude  threshold for the detection
of  the external  modulation and  the width  of the
stochastic-resonance  peak   show  considerable   variation as
the randomness of the underlying small-world network is changed.
} \end{abstract}


\maketitle

\section{Introduction}

Many mathematical  models of  social processes,  inspired by
analogies with  physical  systems,  have  recently  been
formulated in order to describe  a  wide  spectrum  of phenomena
\cite{weid}.  Economic and financial processes, disease
spreading  and information  propagation, and evolution of social
structures, among others, have  been analyzed along such lines.
Weidlich \cite{weid} has proposed a model of  public opinion
formation which, in  the simplest variant, considers that the
individuals can adopt two different  opinions. The opinion of a
given individual may be  influenced by that  of the neighbors,
making it to change with a certain probability.  In the present
work, we  analyze a similar model of social influence and opinion
formation.  In contrast with  previous studies  of  this  class
of  models,  however, we are interested at taking into account
the network of social  interactions. Specifically,  we  model
the  underlying  social  structure of   the population as a
small-world  network.  Small-world networks  \cite{ws}
incorporate two main features of real social interactions. First,
they are highly clustered,  which means that  any two neighbors
of a given site  have  a  relatively large  probability  of being
in turn mutual neighbors. Second, the mean  number of
intermediaries between  any two sites is quite small, and
increases very slowly as the total number of sites grows
\cite{barrat}. This is precisely the small-world property,
originally  discussed  by  Milgram  as  a  typical feature  of
social communities and relationships \cite{milgram}. Small-world
networks can be  considered  as  partially disordered structures.
The construction procedure makes it possible  to control  the
degree  of disorder of a given network, ranging from ordered
lattices to completely disordered graphs. Among many other
applications, these networks have been used as a model of social
structures in the study of disease \cite{ws,MN,KA,SV,ZC} and rumor
propagation \cite{Z}.

As pointed out in the next section, our model is an Ising-like
system whose dynamics is driven by the majority rule.
Equilibrium  properties of Ising systems on small-world networks
have been studied in  detail. It  has  been  shown  that,  in
asymptotically large systems, even an infinitesimal amount of
disorder on a one-dimensional Ising lattice is sufficient for the
system to undergo a ferromagnetic phase  transition as the
temperature varies  \cite{barrat}. Here,  instead, we consider
the  case  where  the  system  is  maintained out of equilibrium
by an external modulation. In the frame of the problem of opinion
formation, this modulation  plays the  role of  a periodic
``fashion'' wave,  as actually  observed  in  certain  real
situations  \cite{babi}.    The combination  of  this modulation
with noise---which, in equilibrium, would  play  the role  of
the  temperature---leads  naturally to the consideration of
stochastic resonance \cite{srrmp}. In the phenomenon  of
stochastic resonance, an enhancement of the response of a system
to an external modulation  is obtained  by an  adequate choice of
the noise level.  This  effect  has attracted  great interest due
to both its potential technological   applications and   its
connection    with biological detection mechanisms. The range of
possible applications in science  includes paleoclimatology
\cite{BEN}, electronic  circuits \cite{elect1,elect2},  lasers
\cite{laser,laser2},  chemical \cite{chem} and biological
systems \cite{biol,biol2}.   Stochastic resonance   in Weidlich's
model of  opinion  formation  has already  been  studied
\cite{babi}. In  the present  work, we analyze this  phenomenon
in  a closely related model of social imitation, incorporating a
small-world network as the underlying structure of social
interactions.

% Figure: histograms
\begin{figure}[tbp]
\centering
\resizebox{.7\columnwidth}{!}{\includegraphics{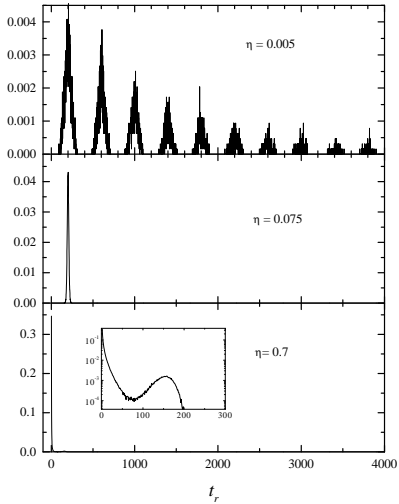}}
\caption{Normalized histograms of the residence time $t_r$, for
$p=0.8$. Three  different noise intensities $\eta$ are
considered. Each histogram has been constructed  recording the
residence times over an interval of $8 \times 10^5$ time units,
with a subsequent average over $20$ realizations of the
small-world network and the initial condition. The insert in the
third plot shows the same data with logarithmic scale in the
vertical axis.} \label{f1}
\end{figure}

\section{The model}

As advanced in the Introduction, we consider an Ising-like model
on  a network with a  controlled degree of  disorder, namely, a
small-world network. The network  is constructed as  follows
\cite{ws}.   We start with  an  $N$-node  one-dimensional
ordered  lattice  with   periodic boundary conditions,  where
each  node is  linked to  its $2K$ nearest neighbors ($K$
clockwise and  $K$ counterclockwise).  The network  is then
disordered,  rewiring each  of the  $K$ clockwise  connections of
each node  $i$ to  a randomly  chosen node  $j$, with probability
$p$. Double and multiple connections are forbidden, and
realizations  where the network becomes  disconnected are
discarded.  The final degree  of disorder, or randomness,  is
determined by  the probability $p$.  As a result of  the
disordering  process, some  shortcuts between otherwise distant
regions are created. Note that, independently of the value  of
$p$, the average number of links per site is always $2K$.

Each site of the small-world network is occupied by an individual
with two possible states, $\mu_i=\pm 1$. At each time step, an
individual $i$ is selected  at random  from the  whole
population,  and the following three evolution rules are
successively applied to its  state $\mu_i$. First, $\mu_i$ is
changed according to the majority rule, i.e.
\begin{equation} \label{sum}
\mu_i \to \mbox{sign}\sum_{j\{i\}} \mu_j ,
\end{equation}
where the  sum runs  over the  neighborhood of  site $i$.   If
the sum equals zero, $\mu_i$ is left unchanged. Second, we apply
the  external modulation. The effect of this modulation is
prescribed by a  harmonic function of time, $f(t)= A \cos (2\pi t
/ \tau)$, with $0<A<1$. Namely, the state of individual $i$
changes according to
\begin{equation}
\mu_i \to \mbox{sign } f(t),
\end{equation}
with  probability  $\Pi_r  (t)=|f(t)|$.  Finally,  noise is
applied as follows. With constant probability  $\Pi_n$ the state
$\mu_i$  is left unchanged. With  the complementary  probability
$1-\Pi_n$,  $\mu_i$ is assigned one of the two possible values
$\pm 1$, chosen at random with probability $1/2$. We call
$\eta=1-\Pi_n$ the noise intensity.

In the absence of modulation, at low noise levels, and for
moderate and large randomness $p$, our population evolves towards
an almost homogeneous state, where $\mu_i$ is the same at
practically all sites. The average of $\mu_i$ over the
population, $\overline \mu=N^{-1} \sum^n _{i=1} \mu_i$, is close
to $\pm 1$. Due to the effect of noise, occasional transitions
between states with opposite signs are observed. For sufficiently
small $p$, on the other hand, $\overline \mu$ fluctuates around
zero with relatively small amplitude. These two regimes
correspond, respectively, to the ferromagnetic and paramagnetic
phases predicted for equilibrium small-world Ising systems
\cite{barrat}. In our simulations, we will mainly consider
randomness and noise levels for which the nonmodulated system is
in the ferromagnetic-like phase. All the cases with $p>0$ will in
fact correspond to that phase. In this situation, the noiseless
version of our system exhibits bistability. The paramagnetic-like
phase will only be encountered in the special case $p=0$.

The effect of the external modulation on the ferromagnetic-like
phase consists  of  a  periodic  modification in the transition
probability between the  two states  with $\overline  \mu
\approx  \pm 1$.  In the paramagnetic-like phase, on the other
hand, $\overline \mu$ oscillates around zero at the same
frequency as the modulation. In both cases, of course, the
modulation is  combined with  the random  fluctuations of noise.
It  is  this  nontrivial  interplay,  which  may result in the
enhancement of the system response, that we focus on in our
numerical simulations.

% Figure: Avsp
\begin{figure}[tbp]
\centering \resizebox{\columnwidth}{!}{\includegraphics{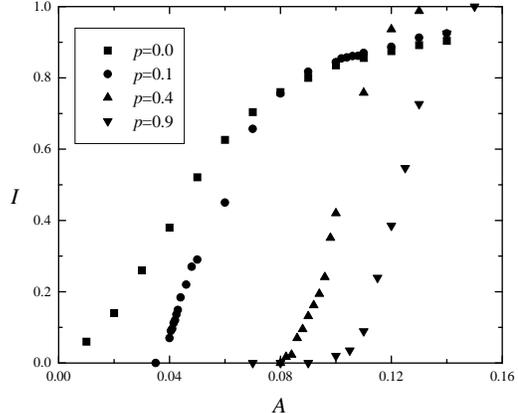}}
\caption{Response $I$ as a function of the modulation amplitude
$A$, for $\eta=0.1$, and four values of the randomness $p$. }
\label{f2}
\end{figure}

\section{Numerical results}

We  have  performed  extensive  numerical  simulations  of  the
model presented in the previous section, considering small-world
networks of $N=10^3$ nodes with $K=3$. The time interval assigned
to an  evolution step is $\Delta t=N^{-1}$ so that,  on the
average, the state of  each site is updated  once during each
time unit. Each  realization starts with  the   generation  of
the  random   network  and   the   random initialization of  the
state  of all  the elements.  The evolution  is analyzed after  a
transient  of $10^5$  time units,  during which  the system is
found to reach a steady large-time behavior. In  particular, we
record the residence times  $t_r$, i. e. the intervals  between
two consecutive changes of sign in the average state $\overline
\mu$. Note that the external modulation is expected to induce
changes in the sign of $\overline \mu$ with a typical period
close to $\tau/2$. The values of $t_r$ are used to  construct a
histogram, showing the  frequency of occurrence of each residence
time. Throughout all the calculations we considered  $t_r=400$.

% Figure: Avsp
\begin{figure}[tbp]
\centering
\resizebox{\columnwidth}{!}{\includegraphics{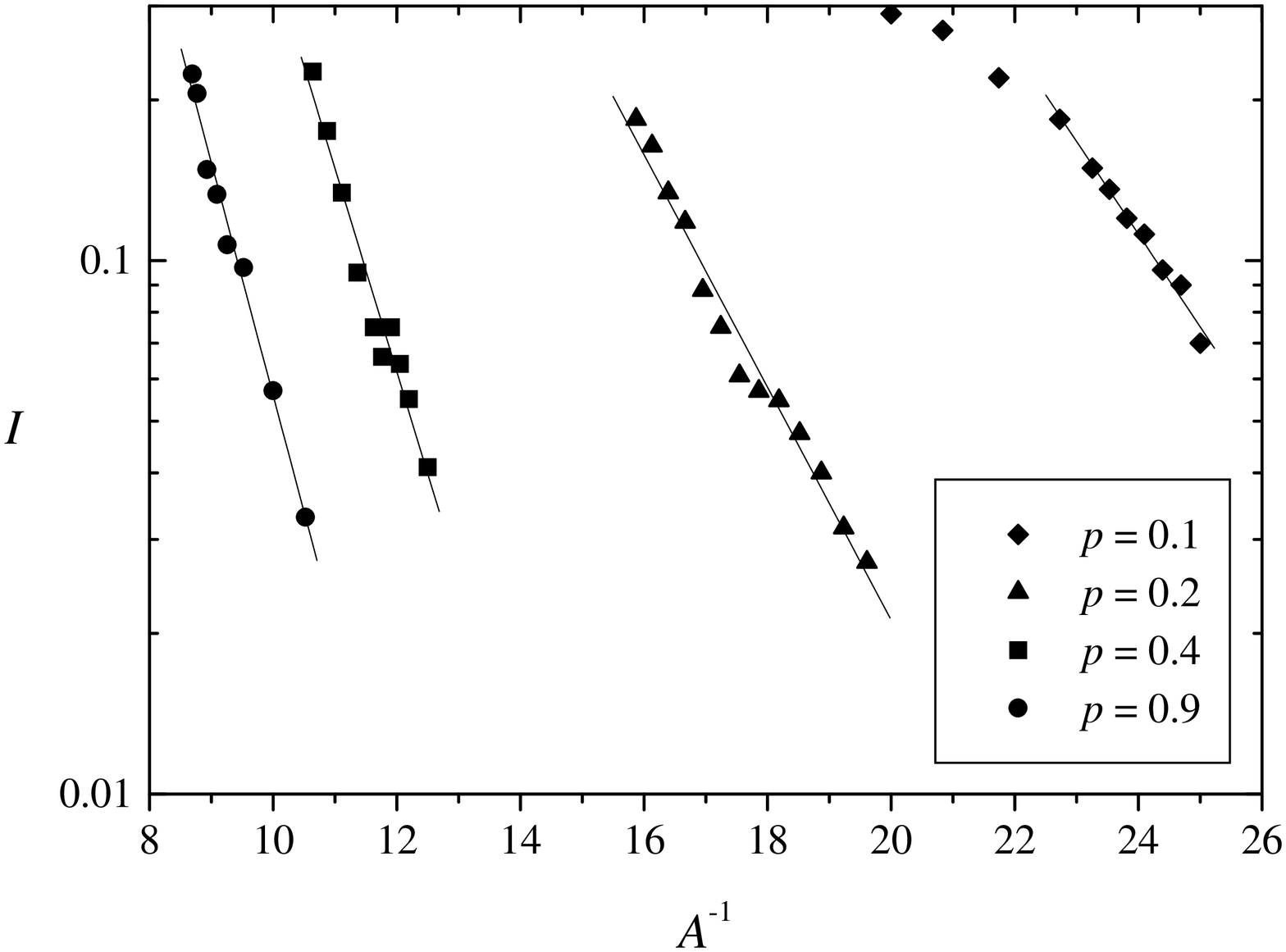}}
\caption{The response $I$, in logarithmic scale, as a function of
the inverse modulation amplitude, $A^{-1}$, for four values of
$p$. Straight lines are least-square fittings to the linear
regime of each data set.} \label{f3}
\end{figure}

Figure \ref{f1} illustrates this  histogram for  fixed modulation
and small-world randomness, and three levels of noise. For low
noise,  the histogram shows a peak at  $t_r=\tau/2$, as expected,
but several  odd harmonics,  $t_r=k\tau/2$  with  $k=3,5,7,\dots$,
are  also observed. These harmonics  correspond to  residence
intervals  during which  the modulation fails, one  or more
times,  at inducing the  transition. At intermediate noise
levels,  corresponding to the  resonance intensity, only  the
pick  around  $t_r=\tau/2$  is  important. For higher noise
levels the transitions are disordered and highly frequent, so
that the residence times are usually very short and the histogram
shows a  peak near $t_r=0$. We stress that  the analysis of these
patterns  has been employed  as  a  quantitative  description  of
stochastic resonance in other systems \cite{srrmp,cast}, as an
alternative to the study of the signal-to-noise ratio
\cite{srrmp}. In fact, a measure of the sensitivity to the
external  modulation is given  by the area  $I$ under the  peak
around $\tau/2$ in the (normalized) histogram. Below, we refer
to  the area $I$ as the {\em response} of the system.

% Figure: I vs. \eta
\begin{figure}[tbp]
  \centering
  \resizebox{\columnwidth}{!}{\includegraphics{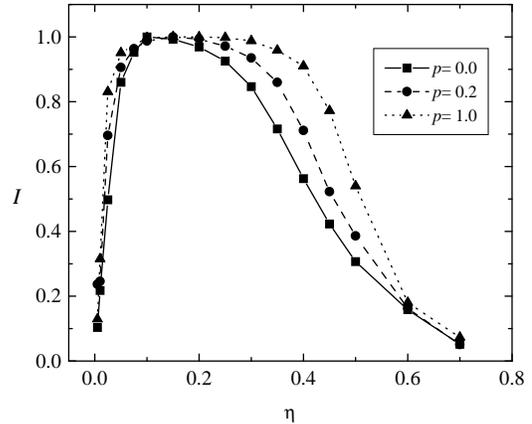}}
\caption{The response $I$ as a function of the noise intensity
$\eta$, for three values of the small-world randomness $p$. Lines
joining dots are drawn as a guide to the eye.} \label{f4}
\end{figure}

First, we analyze the behavior of our system as the modulation
amplitude $A$  is varied.   Fixing the intensity of noise $\eta$
we observe, for $p>0$, the existence of a threshold in $A$ below
which the system  is not able to respond to the modulation. We
find that, as the randomness $p$ increases, the threshold
amplitude grows as well. Figure  \ref{f2} shows the response $I$
for a noise intensity $\eta=0.1$ as a  function of $A$,  for
several  values of  $p$. The nature  of this threshold
phenomenon  is  clarified in Fig. \ref{f3},  where we plot $I$ in
logarithmic scale  as a function of  the inverse amplitude for
three values of  $p$. The approximately  linear dependence  for
small  $I$ suggests the functional form
\begin{equation}
I(A) =I_0 \exp(-A_0/A).
\end{equation}
This  form  is  reminiscent  of  Kramers  formula  for  the
transition frequency  $\omega$  between  equilibrium  states
under the action of thermal  noise  $T$,  $\omega  \propto  \exp
(-T_0/T)$, where $T_0$ is related   to   the   potential
barrier   separating  the  equilibria \cite{Kramers}. In  our
case,  for a  fixed noise  level, the  role of temperature---as
the factor that induces the transition---is played by the
external modulation. Exploiting this analogy, our model can be
thought of, at the macroscopic level, as a system subject to an
effective bistable potential, with a barrier proportional to the
slope $A_0$. It is apparent from Fig. \ref{f3} that the slope
depends on the small-world randomness. As $p$ grows from $0.1$ to
$0.9$, the slope increases by a factor of about $2.5$, indicating
that the minima in the effective potential become deeper for
larger randomness.

Taking into account the results shown in Figs. \ref{f2} and
\ref{f3}, in our study of dependence of the response $I$ on the
noise  intensity $\eta$  we  fix  $A=0.2$  which,  for  the values
of  $p$ and $\eta$ considered  in  our  simulations, is  always
above  the   detection threshold. In Fig. \ref{f4}, we show $I$
as a function of $eta$ for several values of $p$. The existence
of a resonance as the noise intensity is varied is  apparent. At
the  same  time,  an  effect related to the variation of  the
randomness  $p$ is  observed. Namely, the higher is $p$, the
broadest is  the resonant  interval in  the $eta$.  In other
words, accurate  tuning of  the noise  level in order to enhance
the response is less important in more disordered networks.

\begin{figure}[tbp]
  \centering
  \resizebox{\columnwidth}{!}{\includegraphics{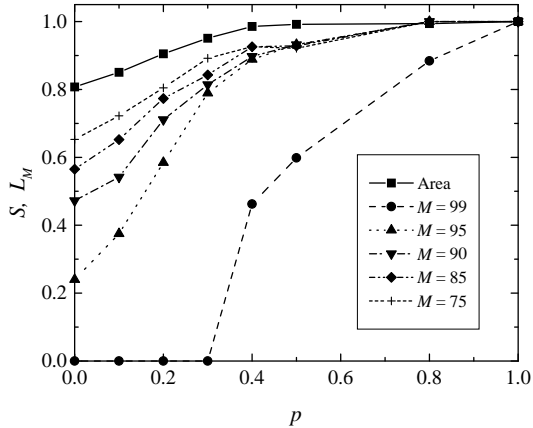}}
\caption{Area $S$ under the stochastic-resonance curves (Fig.
\ref{f4}) and lengths $L_M$ for several values of $M$ (see main
text), as a function of the randomness $p$. Lines joining dots are
drawn as a guide to the eye.} \label{f5}
\end{figure}

Figure \ref{f5} presents a quantitative characterization of the
broadening of the resonance peaks as $p$ grows. First, we show the
area $S$ of the peaks as a function of $p$. Moreover, we show the
length $L_M$ of the intervals of noise intensity where the
response $I$ is above the level $I_M=0.01M$, for several values
of $M$.

% Figure: area vs. p

\section{Conclusion}

We have presented an Ising-like model of opinion formation by
imitation on a small-world network, subject to the action of an
external modulation and noise. The study is focused on the
phenomenon of stochastic resonance, which underlies the
macroscopic behavior of this system. The main control parameters
in our analysis are the amplitude $A$ of the external modulation,
the intensity of noise $\eta$, and the small-world randomness
$p$. The first aspect studied here is the existence of a
threshold in the amplitude of the external signal, below which
the system is not able to detect the modulation. This threshold
is higher as the disorder of the network increases, and
disappears for $p=0$. This fact suggests that the existence of a
threshold in $A$ is directly related to the bistable nature of
the effective potential which acts on the system in its
ferromagnetic-like phase, i. e. for moderate and large
randomness. It is interesting to mention that the increase of the
threshold amplitude with the randomness, which amounts to a
growth in the effective potential barrier, has a well-known
correlate in social imitation processes subject to the action of
noise \cite{damian}. The resistance to collective changes of
opinion grows drastically when long-range social interactions are
allowed. A population restricted to local interactions, instead,
is susceptible to global opinion transitions by diffusive-like
propagation.

As expected, the analysis of the response of the system as a
function of the noise intensity $\eta$---provided that the
amplitude $A$ is above the detection threshold---reveals the
occurrence of a resonance phenomenon. Though the response of the
system at the resonance is practically independent of the
randomness $p$, an enhancement of the phenomenon is apparent from
the broadening of the resonance curves as a function of $\eta$
(Figs. \ref{f4} and \ref{f5}). The convenience of broadening the
stochastic-resonance peak to avoid noise tuning has been
discussed in connection with a variety of systems
\cite{cast,bowi,claudio}.

The existence of stochastic resonance in a model of opinion
formation yields the appealing implication that there is an
optimal noise level for a population to respond to an external
``fashion'' modulation \cite{babi}. Lower noise intensities lead
to the dominance of the majority's opinion, irrespectively of the
external influences, while sufficiently stronger fluctuations
prevent the formation of a definite collective opinion. We have
here shown that, in this phenomenon, the underlying structure of
social interactions plays a role, allowing for a looser tuning of
the resonance noise intensity as the disorder grows.

Let us finally mention that we have tested the robustness of
these results by studying a few variations of our model. For
instance, we have considered the case where the external
modulation acts only when it is not possible to define an
individual's state by the majority rule, i. e. when the sum in Eq.
(\ref{sum}) equals zero. In this case, the individual adopts the
state dictated by the ``fashion'' wave with probability
$\Pi_r(t)$, defined in Sect. II, or preserves the previous state
with the complementary probability. In all cases we got
consistent results, qualitatively similar to those presented
here. Though our model is an oversimplified caricature of any
real process, the results illustrate the complex interplay
between fluctuations, external influences, and interaction
structure that is expected to take place in actual societies.

\begin{acknowledgments}
The authors thank  H. S. Wio for fruitful suggestions. MK thanks
Fundaci\'on Antorchas for financial support.
\end{acknowledgments}

\newpage
\end{document}